\documentclass[runningheads]{llncs}

\usepackage{geometry}
\usepackage[T1]{fontenc}
\usepackage{graphicx}
\usepackage{authblk}
\usepackage{enumerate}
\usepackage[inline]{enumitem}   
\usepackage{mathtools}
\usepackage{algorithm}
\usepackage[noend]{algpseudocode}
\usepackage{algpseudocodex}
\algrenewcommand\algorithmicrequire{\textbf{Input:}}
\algrenewcommand\algorithmicensure{\textbf{Output:}}
\usepackage{eqparbox,array}
\usepackage{xspace}
\usepackage{amsfonts,amssymb}
\usepackage[mathscr]{euscript}
\newcommand{\XX}[1]{\normalfont{\upshape{#1}}}

\usepackage{color}

\newcommand{\rsat}{\emph{Fitch-sat}\xspace}
\newcommand{\Rone}{\ensuremath{\overrightarrow{1}}} 
\newcommand{\Virr}{\ensuremath{R^{\times}_{V}}}
\newcommand{\Wirr}{\ensuremath{R^{\times}_{W}}}
\newcommand{\Cirr}{\ensuremath{R^{\times}_{C}}}

\newcommand{\mcEs}{\ensuremath{\mathcal{E}^{*}}}

\newcommand{\indeg}{\ensuremath{\mathrm{indeg}}} 
\newcommand{\outdeg}{\ensuremath{\mathrm{outdeg}}}

\newcommand{\PROBLEM}[1]{{\sc #1}}

\newcommand{\join}{\mathrel{\ooalign{\hss$\triangleleft$\hss\cr$\triangleright$}}}
\newcommand{\djoin}{\triangleright}

\newcommand{\btwn}{\mathop{::}}   
\newcommand{\noedge}{\mathop{\text{\textvisiblespace}}}
\DeclareMathOperator{\lca}{lca}

\begin{document}
\title{Fitch Graph Completion \thanks{This work was supported in part by
    the German Research Foundation (DFG, STA 850/49-1) and 
    the Data-driven Life Science (DDLS) program funded by
    the Knut and Alice Wallenberg Foundation}}
%
%
\author{Marc Hellmuth\inst{1}\orcidID{0000-0002-1620-5508} \and
Peter F.\ Stadler\inst{2}\orcidID{0000-0002-5016-5191}
\and Sandhya Thekkumpadan Puthiyaveedu\inst{1}\orcidID{0000-0002-7745-3935} }
\authorrunning{Hellmuth et al.}
%
\institute{Dept.\ Mathematics, Faculty of Science,
  Stockholm University, SE-10691 Stockholm, Sweden.
  \email{marc.hellmuth@math.su.se}, \email{thekkumpadan@math.su.se}
   \and
    Bioinformatics Group, Dept.\ Computer Science \&
    Interdisciplinary Center for Bioinformatics, Universit{\"a}t Leipzig,
    D-04107 Leipzig, Germany. \email{studla@bioinf.uni-leipzig.de}
 }
\maketitle              
\begin{abstract}
  Horizontal gene transfer is an important contributor to evolution.
  According to Walter M.\ Fitch, two genes are xenologs if they are
  separated by at least one HGT. More formally, the directed Fitch graph
  has a set of genes is its vertices, and directed edges $(x,y)$ for all
  pairs of genes $x$ and $y$ for which $y$ has been horizontally
  transferred at least once since it diverged from the last common ancestor
  of $x$ and $y$. Subgraphs of Fitch graphs can be inferred by comparative
  sequence analysis. In many cases, however, only partial knowledge about
  the ``full'' Fitch graph can be obtained. Here, we characterize
  Fitch-satisfiable graphs that can be extended to a biologically feasible
  ``full'' Fitch graph and derive a simple polynomial-time recognition
  algorithm. We then proceed to showing that finding the Fitch graphs with
  total maximum (confidence) edge-weights is an NP-hard problem.


\end{abstract}
\sloppy

\section{Introduction}

Horizontal gene transfer (HGT) is a biological process by which genes from
sources other than the parents are transferred into an organisms genome. In
particular in microorganism it is an important contributor to evolutionary
innovation.  The identification of HGT events from genomic data, however,
is still a difficult problem in computational biology, both mathematically
and in terms of practical applications. Recent formal results are phrased
in terms of a binary relation between genes. In most situations it can be
assumed that the evolution of genes is tree-like and thus can described by
a gene tree $T$ whose leaves correspond to the present-day, observable
genes; the interior vertices and edges then model evolutionary events such
as gene duplications, speciations, and also HGT. Since HGT
distinguishes between the gene copy that continues to be transmitted
vertically, and the transferred copy, one associated the transfer with the
edge in $T$ connecting the HGT event with its transferred
offspring. Focusing on a pair of present-day genes, it is of interest to
determine whether or not they have been separated by HGT events in their
history. This information is captured by the Fitch (xenology) graph. It
contains an edge $x\to y$ whenever a HGT event occurred between $y$ and the
least common ancestor of $x$ and $y$ \cite{Geiss:18a}.  Fitch graphs form a
hereditary sub-class of the directed cographs \cite{CP-06}, and admit a
simple characterization in terms of eight forbidden induced subgraphs on
three vertices (see Fig.~\ref{fig:forbidden} below). Moreover, every Fitch
graph uniquely determines a least resolved edge-labeled tree by which it is
explained.  This tree is related to the gene tree by a loss of resolution
\cite{Geiss:18a}.

Information on HGT events can be extracted from sequence information using
a broad array of methods \cite{Ravenhall:15}, none of which is likely to
yield a complete picture.  Reliable information to decide whether or
not two genes are xenologs, thus, may be available only for some pairs of
genes $(x,y)$, but not for others.  In this situation it is natural to ask
whether this partial knowledge can be used to infer missing information. In
\cite{NEMWH:18} the analogous question was investigated for di-cographs.
The main result of the present contribution is a characterization of
partial Fitch graphs, Thm.~\ref{thm:Characterize-Partial}, and an
accompanying polynomial-time algorithm. In addition, we show that the
``weighted'' version of Fitch graph completion is NP-hard.

\section{Preliminaries}
  
{\textbf{Relations.}}
Throughout, we consider only \emph{irreflexive}, \emph{binary} relations
$R$ on $V$, i.e., $(x,y)\in R$ implies $x\ne y$ for all $x,y\in V$. We
write $\overleftarrow{R} \coloneqq \{(x,y) \mid (y,x)\in R\}$ and
$R^{\mathrm{sym}}\coloneqq R\cup\overleftarrow{R}$ for the \emph{transpose}
and the \emph{symmetric extension} of $R$, respectively. The relation
$R^{\times}_V \coloneqq\{(x,y)\mid x,y\in V,\,x\ne y\}$ is called the
\emph{full} relation.  For a subset $W\subseteq V$ and a relation $R$, we
define the \emph{induced} subrelation as $R[W] \coloneqq \{(x,y)\mid
(x,y)\in R, x,y\in W\}$.  Moreover, we consider ordered tuples of relation
$\mathcal{R}=(R_1,\dots,R_n)$.  Let $R_1,\dots,R_n\subseteq \Virr$, then
$\mathcal{R}=(R_1,\dots,R_n)$ is \emph{full} if $\cup_{i=1}^n R_i = \Virr$
and \emph{partial} if $\cup_{i=1}^n R_i \subseteq \Virr$. Note that a full
tuple of relations is also considered to be a partial one. Moreover, we
consider component-wise sub-relation and write $\mathcal{R}[W]\coloneqq
(R_1[W],\dots, R_n[W])$ and $\mathcal{R} = (R_1,\dots,R_n) \subseteq
\mathcal{R}' = (R'_1,\dots,R'_n)$ if $R_i\subseteq R'_i$ holds for all
$i\in\{1,\dots,n\}$.  In the latter case, we say that $\mathcal{R'}$
\emph{extends} $\mathcal{R}$.

\medskip\noindent
{\textbf{Digraphs and DAGs.}}
A directed graph (digraph) $G=(V,E)$ comprises a vertex set $V(G)=V$ and an
irreflexive binary relation $E(G)=E$ on $V$ called the edge-set of $G$.
Given two disjoint digraphs $G=(V,E)$ and $H=(W,F)$, the digraphs $G\cup H
= (V\cup W, E\cup F)$, $G\join H = (V\cup W, E\cup F\cup \{(x,y),(y,x)\mid
x\in V,y\in W\})$ and $G\djoin H = (V\cup W, E\cup F\cup \{(x,y)\mid x\in
V,y\in W\})$ denote the \emph{union}, \emph{join} and \emph{directed join}
of $G$ and $H$, respectively.  For a given subset $W\subseteq V$, the
\emph{induced subgraph} $G[W]=(W,F)$ of $G=(V,E)$ is the subgraph for which
$x,y\in W$ and $(x,y)\in E$ implies that $(x,y)\in F$. We call $W\subseteq
V$ a \emph{(strongly) connected component} of $G=(V,E)$ if $G[W]$ is an
\emph{inclusion-maximal} (strongly) connected subgraph of $G$.

Given a digraph $G=(V,E)$ and a partition $\{V_1,V_2,\dots,V_k\}$, $k\geq
1$ of its vertex set $V$, the \emph{quotient digraph
$G/\{V_1,V_2,\dots,V_k\}$ has as vertex set} $\{V_1,V_2,\dots,V_k\}$ and
two distinct vertices $V_i$ and $V_j$ form an edge $(V_i, V_j)$ in
$G/\{V_1,\dots,V_k\}$ if there are vertices $x\in V_i$ and $y\in V_j$ with
$(x,y)\in E$. Note, that edges $(V_i, V_j)$ in $G/\{V_1,\dots,V_k\}$ do not
necessarily imply that $(x,y)$ form an edge in $G$ for $x\in V_i$ and $y\in
V_j$.  Nevertheless, at least one such edge $(x,y)$ with $x\in V_i$ and
$y\in V_j$ must exist in $G$ given that $(V_i, V_j)$ is an edge in
$G/\{V_1,\dots,V_k\}$

A cycle $C$ in a digraph $G=(V,E)$ of length $n$ is an ordered sequence of
$n>1$ (not necessarily distinct) vertices $(v_1,\dots,v_n)$ such that
$(v_n,v_{1}) \in E$ and $(v_i,v_{i+1}) \in E$, $1\leq i<n$.  A digraph that
does contain cycles is a \emph{DAG} (directed acyclic graph). Define the
relation $\preceq_G$ of $V$ such that $v \preceq_G w$ if there is directed
path from $w$ to $v$. A vertex $x$ is a \emph{parent} of $y$ if $(x,y)\in
E$. In this case $y$ is \emph{child} of $x$.  Then $G$ is DAG if and only
if $\preceq_G$ is a partial order. We write $y \prec_T x$ if $y \preceq_T
x$ and $x\neq y$.  A \emph{topological order} of $G$ is a total order $\ll$
on $V$ such that $(v,w)\in E$ implies that $v\ll w$.
It is well known
that a digraph $G$ admits a topological order if and only if $G$ is a
DAG. In this case, $x\prec_G y$ implies $y \ll x$, i.e., $\ll$ is a linear
extension of $\prec$. Note that $\preceq$ and $\ll$ are arranged in
opposite order.  The effort to check whether $G$
admits a topological order $\ll$ and, if so, to compute $\ll$ is linear,
i.e., in $O(|V|+|E|)$ \cite{Kahn:62}. If $C_1,\dots,C_k$, $k\geq 1$ are the
strongly connected components of a directed graph, then
$G/\{C_1,C_2,\dots,C_k\}$ is a DAG.

A DAG $G$ is \emph{rooted} if it contains a unique $\preceq_G$-maximal
element $\rho_G$ called the \emph{root}. Note that $\rho_G$ is
$\ll$-minimal. A rooted tree $T$ with vertex set $V(T)$
is a DAG such that every vertex $x\in V(T)\setminus\{\rho_T\}$ has a unique
parent. The rooted trees considered here do not contain vertices $v$ with
$\indeg(v) = \outdeg(v)=1$.  A vertex $x$ is an \emph{ancestor} of $y$ if
$x\preceq_T y$, i.e., if $x$ is located on the unique path from $\rho_T$ to
$y$.  A vertex in $T$ without a child is a \emph{leaf}. The set of leaves
of $T$ will be denoted by $L(T)$. The elements in $V^0(T)\coloneqq V(T)\setminus
L(T)$ are called the inner vertices.  We write $T(u)$ for the subtree of
$T$ induced by $\{v\in V(T)\mid v\preceq_T u\}$. Note that $u$ is the root of
$T(u)$.

For a subset $W\subseteq L(T)$, the \emph{least common ancestor}
$\lca_T(W)$ of $W$ is the unique $\preceq_T$-minimal vertex that is an
ancestor of each $w\in W$.  If $W=\{x,y\}$, we write $\lca_T(x,y) \coloneqq
\lca_T(\{x,y\})$. A rooted tree $T$ is \emph{ordered}, if the children of
every vertex in $T$ are ordered.

We say that rooted trees $T_1, \dots, T_k$, $k\geq 2$ \emph{are joined under a
new root in the tree $T$} if $T$ is obtained by the following procedure: add a
new root $\rho_T$ and all trees $T_1, \dots, T_k$ to $T$ and connect the root
$\rho_{T_i}$ of each tree $T_i$ to $\rho_T$ with an edge $(\rho_T,\rho_{T_i})$.

\medskip\noindent
{\textbf{Directed Cographs.}}
Di-cographs generalize the notion of undirected cographs
\cite{EHPR:96,CP-06,Corneil:81,Corneil:85} and are defined recursively as
follows: (i) the single vertex graph $K_1$ is a di-cograph, and (ii) if $G$ and
$H$ are di-cographs, then $G\cup H$, $G\join H$, and $G\djoin H$ are
di-cographs \cite{gurski2017dynamic,NEMWH:18}. Every di-cograph $G=(V,E)$
\emph{is explained by} an ordered rooted tree $T=(W,F)$, called a
\emph{cotree} of $G$, with leaf set $L(T)=V$ and a labeling function
$t:W^0\to \{0,1,\overrightarrow{1}\}$ that uniquely determines the
set of edges $E(G) = E_{1}(T,t)\cup E_{\Rone}(T,t)$ and set of non-edges
$E_{0}(T,t)$ of $G$ as follows:
\begin{align*}
  & E_{1}(T,t) = \{(x,y)\mid t(\lca(x,y)) =1\}, \\ 
  & E_{0}(T,t) = \{(x,y)\mid  t(\lca(x,y)) =0\}, \text{ and}\\ 
  & E_{\Rone}(T,t) = \{(x,y)\mid t(\lca(x,y))=\Rone \text{ and } x
  \text{ is left of } y \text{ in } T\}. 
\end{align*}
Note that $E_{i}(T,t) = E_{i}(T,t)^{\mathrm{sym}}$ for $i\in \{0,1\}$ since
$\lca(x,y)=\lca(y,x)$. Every di-cograph $G=(V,E)$ is explained by a unique
\emph{discriminating} cotree $(T,t)$ satisfying $t(x)\neq t(y)$ for all
$(x,y)\in E(T)$.  Every cotree $(T',t')$ that explains $G$ is
``refinement'' of it discriminating cotree $(T,t)$, i.e., $(T,t)$ is
obtained by contracting all edges $(x,y)\in E(T')$ with $t'(x)=t'(y)$
\cite{Boecker:98}.  Determining whether a digraph is a di-cograph, and if
so, computing its discriminating cotree requires $O(|V|+|E|)$ time
\cite{Corneil:81,gurski2017dynamic,mcconnell2005linear}.

\section{Fitch graphs and Fitch-satisfiability}

\paragraph{\textbf{Basic properties of Fitch graphs.}}
Fitch Graphs are defined in terms of edge-labeled rooted trees $T$ with an
\emph{edge-labeling} $\lambda\colon E\to\{0,1\}$ and leaf set $L(T)=V$. The
graph $\mathbb{G}(T,\lambda) = (V,E)$ contains an edge $(x,y)$ for $x,y\in
V$ if and only if the (unique) path from $\lca_T(x,y)$ to $y$ contains at
least one edge $e\in E(T)$ with label $\lambda(e)=1$. The edge set of
$\mathbb G(T,\lambda)$ by construction is a binary irreflexive relation on
$V$.  A directed graph $G$ is a \emph{Fitch graph} if there is a tree
$(T,\lambda)$ such that $G \simeq \mathbb{G}(T,\lambda)$. Fitch graphs form
a proper subset of directed cographs \cite{Geiss:18a}. Therefore, they can
be explained by a cotree.

\begin{figure}[t]
  \centering
  \includegraphics[width = 0.7\textwidth]{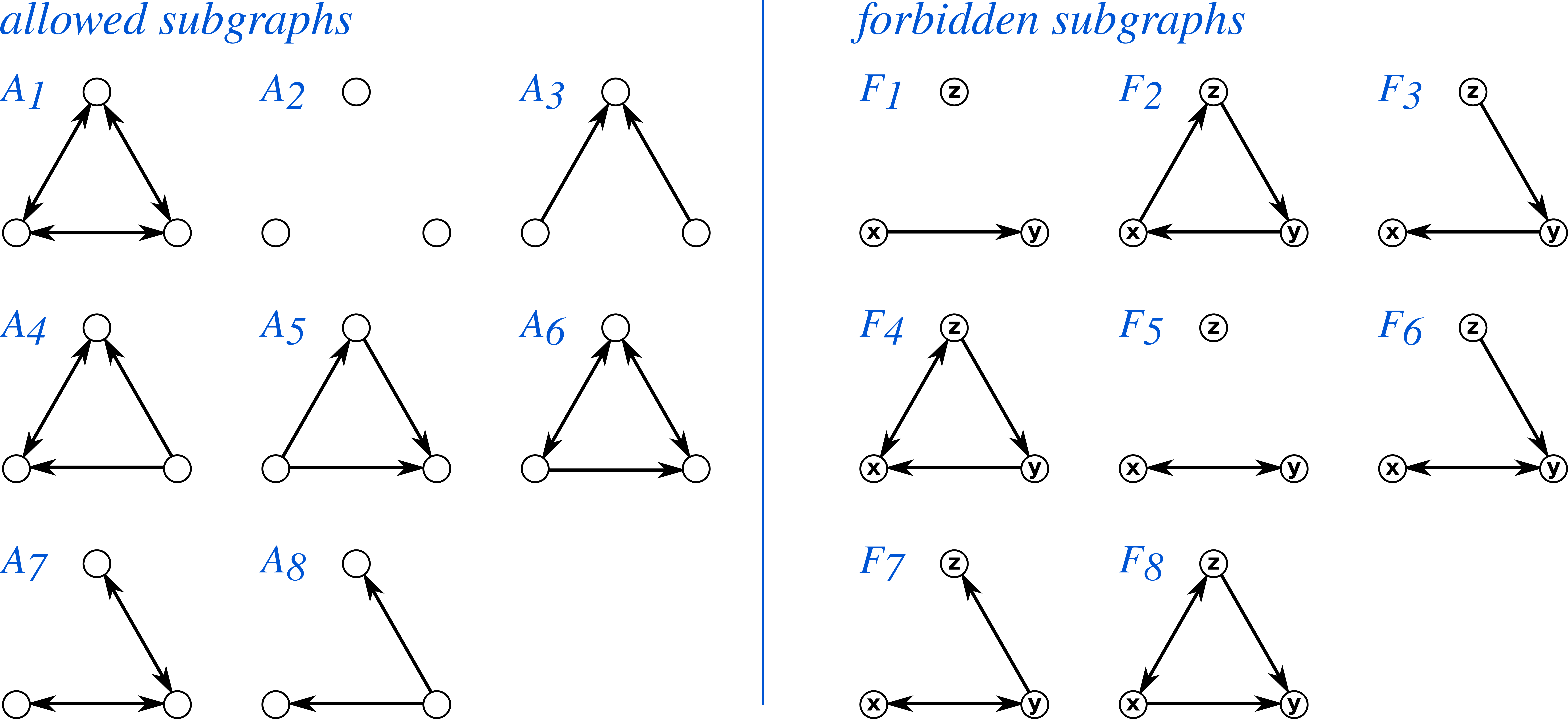}
  \caption{Of the 16 possible irreflexive binary relations on three
    vertices, eight ($A_1$ through $A_8$) may appear in Fitch graphs, while
    the remaining eight ($F_1$ through $F_8$) form forbidden induced
    subgraphs.   
  }
\label{fig:forbidden}
\end{figure}

\begin{definition}
  A cotree $(T,t)$ is a \emph{Fitch-cotree} if there are no two vertices
  $v,w\in V^0(T)$ with $w\prec_T v$ such that either (i) $t(v) = 0 \neq
  t(w)$ or (ii) $t(v) = \Rone$, $t(w) = 1$, and $w\in V(T(u))$ where $u$ is
  a child  of $v$ distinct from the right-most child of $v$.
\end{definition} 
\noindent
In other words, a Fitch-cotree satisfies:
\begin{itemize}
\item[(a)] No vertex with label $0$ has a descendant with label $1$ or $\Rone$.
\item[(b)] If a vertex $v$ has label $\Rone$, then the subtree $T(u)$
  rooted at a $u$ child of $v$ --- except the right-most one --- does not
  contain vertices with label $1$. In particular, if $T$ is discriminating,
  then $T(u)$ is either a star-tree whose root $u$ has label $t(u)=0$ or
  $u$ is a leaf. In either case, the di-cograph $G[L(T(u))]$ defined by the
  subtree $T(u_1)$ of left-most child $u_1$ of $v$, is edge-less.
\end{itemize}		

Fitch graphs have several characterizations that will be relevant
throughout this contribution. We summarize
\cite[L.\ 2.1]{hellmuth2023resolving} and \cite[Thm.\ 2]{Geiss:18a}
in the following 
\begin{theorem}\label{thm:CharFitch}
  For every digraph $G=(V,E)$ the following statements are equivalent. 
  \begin{enumerate}
  \item $G$ is a Fitch graph.
  \item $G$ does not contain an induced $F_1, F_2,\dots, F_8$
    (cf.\ Fig.\ \ref{fig:forbidden}).
  \item $G$ is a di-cograph that does not contain an induced $F_1$, $F_5$ and
    $F_8$ (cf.\ Fig.\ \ref{fig:forbidden}).
  \item $G$ is a di-cograph that is explained by a Fitch-cotree.
  \item Every induced subgraph of $G$ is a Fitch graph, i.e.,
    the property of being a Fitch graph is hereditary. 
  \end{enumerate}
  Fitch graphs can be recognized in $O(|V|+|E|)$ time.  In the affirmative
  case, the (unique least-resolved) edge-labeled tree $(T,\lambda)$ can be
  computed in $O(|V|)$ time.
\end{theorem}
Alternative characterizations can be found in \cite{HS:19}.
The procedure \texttt{cotree2fitchtree} described in \cite{Geiss:18a}
can be used to transform a Fitch cotree $(T,t)$ that explains a Fitch graph
$G$ into an edge-labeled tree $(T',\lambda)$ that explains $G$ in
$O(|V(T)|)$ time, avoiding the construction of the di-cograph
  altogether.
  For later reference, we provide the following simple results.

\begin{lemma}\label{lem:acyclicFitch}
  The graph obtained from a Fitch graph by removing all bi-directional
  edges is a DAG.
\end{lemma}
\begin{proof}
  Let $G = (V,E)$ be the graph obtained from the Fitch graph $F$ by
  removing all bi-directional edges and assume, for contradiction, that $G$
  is not acyclic. Hence, there is a smallest cycle $C = (v_1,\dots,v_n)$ in
  $G$ with $n\geq 3$ vertices.  If $n=3$, then $G$ contains $F_2$ as an
  induced subgraph and thus, $F_2$ is an induced subgraph of $F$. This
  contradicts Thm.\ \ref{thm:CharFitch} and thus the assumption that $F$ is
  a Fitch graph. Hence, $n>3$ must hold. Now consider $v_1,v_2$ and $v_3$
  on $C$ with edges $(v_1,v_2), (v_2,v_3)\in E$. Since $G$ does not contain
  cycles on three vertices, we have $(v_3,v_1)\notin E$ and either
  $(v_1,v_3)\in E$ or $(v_1,v_3)\notin E$.  If $(v_1,v_3)\in E$, then $G$
  contains a smaller cycle $(v_1,v_3,\dots,v_n)$; a contradiction to the
  choice of $C$ as a shortest cycle. Hence, $(v_1,v_3)\notin E$ must
  hold. This leaves two cases for $F$:
\begin{enumerate} 
\item $(v_1,v_3), (v_3,v_1)\in E(F)$. In this case $v_1$, $v_2$ and $v_3$
  induce an $F_4$ in $F$;
\item $(v_1,v_3), (v_3,v_1)\notin E(F)$, i.e.,
  $v_1$ and $v_3$ are not adjacent in $F$. In this case 
  $v_1$, $v_2$ and $v_3$ induce an $F_3$ in $F$. 
\end{enumerate}
Both alternatives contradict the assumption that $F$ is a Fitch graph.
Thus $G$ must be acyclic.
\hfill\qed\end{proof}
\begin{corollary}\label{cor:acyclicFitch}
  Every Fitch graph without bi-directional edges is a DAG.
\end{corollary}
Removal of the bi-directional edges from the Fitch graph $A_6$ yields the
graph $F_1$, i.e., although removal of all bi-directional edges from Fitch
graphs yields a DAG it does not necessarily result in a Fitch graph.
\begin{corollary}\label{cor:DAG-fitch}
  Let $G$ be a directed graph without non-adjacent pairs of vertices and no
  bi-directional edges.  Then, $G$ is a Fitch graph if and only if it is a
  $DAG$.
\end{corollary}
\begin{proof}
  Suppose that $G$ is a directed graph without non-adjacent pairs of
  vertices and no bi-directional edges.  Hence, $G$ cannot contain the
  forbidden subgraphs $F_1$ and $F_3,\dots, F_8$.  If $G$ is a DAG, it also
  cannot contain $F_2$. Thus Thm.\ \ref{thm:CharFitch} implies that $G$ a
  Fitch graph. Conversely, if $G$ is a Fitch graph without bidirectional
  edges, then Cor.~\ref{cor:acyclicFitch} implies that $G$ is a DAG.
\hfill\qed\end{proof}

\paragraph{\textbf{Characterizing Fitch-satisfiability.}} 
Throughout we consider 3-tuples of (partial) relations $\mathcal{E}
=(E_{0},E_{1},E_{\Rone})$ on $V$ such that $E_{0}$ and $E_{1}$ are
symmetric and $E_{\Rone}$ is antisymmetric.
\begin{definition}[\rsat]\label{def:rsat}
  $\mathcal{E}$ is \emph{Fitch}-satisfiable (in short \rsat), if there is a
  full tuple $\mcEs=\{E^*_{0},E^*_{1},E^*_{\Rone}\}$ (with $E^*_{0}$ and
  $E^*_{1}$ being symmetric and $E^*_{\Rone}$ being antisymmetric) that
  extends $\mathcal{E}$ and that is explained by a Fitch-cotree $(T,t)$.
\end{definition}

By slight abuse of notation, we also say that, in the latter case,
$\mathcal E$ is explained by $(T,t)$. The problem of finding a tuple
$\mcEs$ that extends $\mathcal E$ and that is explained by an arbitrary
cotree was investigated in \cite{NEMWH:18}.  Thm.\ \ref{thm:CharFitch}
together with the definition of cotrees and Def.\ \ref{def:rsat} implies
\begin{corollary}\label{cor:EquiDefSat}
  $\mathcal{E} = (E_0,E_1,E_{\Rone})$ on $V$ is \rsat precisely if it can
  be extended to a full tuple $\mcEs =(E^*_0,E^*_1,E^*_{\Rone})$ for which
  $H=(V,E^*_1\cup E^*_{\Rone})$ is a Fitch graph.  In particular, there is
  a discriminating Fitch-cotree that explains $\mathcal{E}$, $\mcEs$,
  and $H$.
\end{corollary}

Fitch-satisfiability is a hereditary graph property.
\begin{lemma}\label{lem:hereditary}
  A partial tuple $\mathcal{E}$ on $V$ is \rsat if and only if
  $\mathcal{E}[W]$ is \rsat for all $W\subseteq V$.
\end{lemma}
\begin{proof}
  The \emph{if} direction follows from the fact that $\mathcal{E}[W]$ is
  \rsat for $W=V$ because $\mathcal{E}[V] = \mathcal{E}$.  For the
  \emph{only-if} direction let $W\subseteq V$ and assume that $\mathcal{E}
  = (E_0,E_1,E_{\Rone})$ is \rsat. By Cor.~\ref{cor:EquiDefSat}),
  $\mathcal{E}$ can be extended to a full 3-tuple $\mcEs
  =(E^*_0,E^*_1,E^*_{\Rone})$ for which $(V,E^*_1\cup E^*_{\Rone})$ is a
  Fitch graph. Since the property of being a Fitch graph is a hereditary
  (Thm.\ \ref{thm:CharFitch}), $(W,E^*_1[W]\cup E^*_{\Rone}[W])$ is again a
  Fitch graph. Moreover, we have $E^*_0[W] = \Wirr \setminus (E^*_1[W]\cup
  E^*_{\Rone}[W]))$, $E_0[W]\subseteq E^*_0[W]$, $E_1[W]\subseteq
  E^*_1[W]$, and $E_{\Rone}[W]\subseteq E^*_{\Rone}[W]$, and thus
  $\mathcal{E}^*[W]=(E^*_0[W],E^*_1[W],E^*_{\Rone}[W])$ is a full tuple
  that extends $\mathcal{E}[W]$.  In summary, $\mathcal{E}[W]$ is \rsat.
\hfill\qed\end{proof}

For the proof of Theorem \ref{thm:Characterize-Partial}, we will need the following technical result.
\begin{lemma}\label{lem:0=edge-less}
  Let $\mathcal{E} = (E_0,E_1,E_{\Rone})$ be a \rsat partial tuple on $V$
  that it is explained by the discriminating Fitch-cotree $(T,t)$ and put
  $G_0[W] \coloneqq (W,E_1[W]\cup E_{\Rone}[W])$ for $W\subseteq V$.  If
  there is a vertex $u\in V^0(T)$ such that $t(u)=0$, then $G_0[C]$ is
  edge-less for all $C\subseteq L(T(u))$.
\end{lemma}
\begin{proof}
  Suppose that $\mathcal{E} = (E_0,E_1,E_{\Rone})$ on $V$ is \rsat and let
  $\mcEs= (E_0^*,E_1^*,E_{\Rone}^*)$ be the full tuple that is explained by
  the discriminating Fitch-cotree $(T,t)$ and that extends $\mathcal{E}$.
  Assume that there is a vertex $u\in V^0(T)$ such that $t(u)=0$. Put
  $W\coloneqq L(T(u))$.  Since $(T,t)$ is a Fitch-cotree, there is no inner
  vertex $v$ such that $v\prec_T u$ and $t(v)\neq 0$. Therefore, and because
  $(T,t)$ is discriminating, all vertices $v\prec_T u$ must be leaves.
  This implies that for all distinct $x,y\in W$ it holds that $\lca(x,y)=u$
  and, thus, $t(\lca(x,y))=t(u) = 0$. Since $(T,t)$ explains
  $\mathcal{E}^*$ it follows that $E_0^*[W] = \Wirr$ and, thus, $E_1^*[W] =
  E_{\Rone}^*[W]=\emptyset$. Since $E_1[W]\subseteq E_1^*[W]$ and
  $E_{\Rone}[W]\subseteq E^*_{\Rone}[W]$, we have $E_1[W] =
  E_{\Rone}[W]=\emptyset$. Hence, $G_0[W] = (W,E_1[W]\cup E_{\Rone}[W])$ is
  edge-less. Trivially, $G_0[C]$ must be edge-less for all $C\subseteq W$.
\hfill\qed\end{proof}

We are now in the position to provide a characterization of \rsat partial
tuples. 
\begin{theorem}\label{thm:Characterize-Partial}
  The partial tuple $\mathcal{E} = (E_0,E_1,E_{\Rone})$ on $V$ is \rsat if
  and only if at least one of the following statements hold.
  \begin{description}
  \item[\XX{(S1)}] $G_0 \coloneqq (V,E_1\cup E_{\Rone})$ is edge-less.
  \item[\XX{(S2)}] \XX{(a)} $G_1 \coloneqq (V,E_0\cup E_{\Rone})$ is
    disconnected and
    \par
    \ \ \XX{(b)}
    $\mathcal{E}[C]$ is \rsat for all connected components $C$ of $G_1$
  \item[\XX{(S3)}]
    \begin{description}
    \item[\XX{(a)(I)}] $G_{\Rone} \coloneqq (V,E_0\cup E_1\cup E_{\Rone})$
      contains $k>1$ strongly connected 
      
      \hspace{0.6cm} components $C_1, \dots, C_k$
      collected in $\mathcal{C}$ and
    \item[\quad\ \ \XX{(II)}] there is a $C\in \mathcal{C}$ for which the
      following conditions are satisfied:
      \begin{description}
      \item[\qquad\XX{(i)}] $G_0[C]$ is edge-less.
      \item[\qquad\XX{(ii)}] $C$ is $\ll$-minimal for some topological
        order $\ll$ 
        \par
        \hspace{0.9cm}on $G_{\Rone} /\{C_1,C_2, \dots, C_k\}$.
      \end{description}
    \item[\ \ \XX{(b)}]
      $\mathcal{E}[V\setminus C]$ is
        \rsat.
    \end{description}	
  \end{description} 
\end{theorem}

\begin{proof}
  We start with proving the \emph{if} direction and thus assume that
  $\mathcal{E} = (E_0,E_1,E_{\Rone})$ is a partial tuple on $V$ that
  satisfies at least one of (S1), (S2) and (S3).

  First assume that $\mathcal{E}$ satisfies (S1). Hence, $G_0 \coloneqq
  (V,E_1\cup E_{\Rone})$ is edge-less and we can replace $E_0$ by
  $E_0^*\coloneqq \Virr$ to obtain the full tuple $\mathcal{E}^* =
  (E_0^*,E_1,E_{\Rone}) = (\Virr,\emptyset,\emptyset)$.  One easily
  verifies that the star tree with leaf-set $V$ and whose root is labeled
  ``0'' explains $\mathcal{E}^*$ and is, in particular, a
  Fitch-cotree. Hence, $\mathcal{E}^*$ is \rsat.

  Next assume that $\mathcal{E}$ satisfies (S2). Hence, $G_1 \coloneqq
  (V,E_0\cup E_{\Rone})$ is disconnected and $\mathcal{E}[C_i]$ is \rsat
  for each connected component $C_1,\dots,C_k$, $k>1$. Thus, we can extend
  each $\mathcal{E}[C_i]$ to a full sat $\mathcal{E}^*[C_i]$ that is
  explained by a Fitch-cotree $(T_i,t_i)$.  Let $\mathcal{E}' =
  (E^*_0,E'_1,E^*_{\Rone})$ be the resulting partial tuple obtained from
  $\mathcal{E}$ such that $E^*_0 = E_0 \cup (\cup_{i=1}^k E^*_0[C_i])$,
  $E'_1 = E_1 \cup (\cup_{i=1}^k E^*_1[C_i])$, and $E^*_{\Rone} = E_{\Rone}
  \cup (\cup_{i=1}^k E^*_{\Rone}[C_i])$. To obtain $E^*_0$ and
  $E^*_{\Rone}$, we added only pairs $(x,y)$ with $x,y\in C_i$ to $E_0$ and
  $E_{\Rone}$, respectively.  Therefore, $G^*_1 \coloneqq (V,E^*_0\cup
  E^*_{\Rone})$ remains disconnected and has, in particular, connected
  component $C_1,\dots,C_k$. We now extend $\mathcal{E}'$ to a full tuple
  $\mathcal{E}^*$ by adding all $(x,y)\in \Virr \setminus (E^*_0\cup
  E'_1\cup E^{*,\mathrm{sym}}_{\Rone})$ to $E'_1$ to obtain $E^*_1$. Note
  that the pairs in $E^*_1$ do not alter disconnectedness of $G^*_1$, i.e.,
  $G^*_1$ has still connected components $C_1,\dots C_k$.  By construction,
  we have $(x,y)\in E^*_1\setminus (\cup_{i=1}^k E^*_1[W])$ with
  $W=\cup_{i=1}^k C_i$ if and only if $x$ and $y$ are in distinct connected
  components of $G^*_1$. We know join the Fitch-cotrees $(T_1,t_1),\dots
  (T_k,t_k)$ under a common root $\rho$ with label ``1'' and keep the
  labels of all other vertices to obtain the cotree $(T,t)$. Since each
  $(T_i,t_i)$ is a Fitch-cotree and since no constraints are imposed on
  vertices with label label ``1'' (and thus on the root $\rho$ of $T$), it
  follows that $(T,t)$ is a Fitch-cotree. Moreover, by construction, for
  all $x$ and $y$ that are in distinct connected components of $G^*_1$ we
  have $\lca_T(x,y)=\rho$ and thus $t(\lca_T(x,y))=1$. Hence, $(x,y)\in
  E^*_1\setminus (\cup_{i=1}^k E^*_1[W]$ if and only of $\lca_T(x,y)=\rho$
  and thus $t(\lca_T(x,y))=1$.  Moreover, each Fitch-cotree $(T_i,t_i)$
  explains $\mathcal{E}^*[C_i]$. Taken the latter arguments together, the
  Fitch-cotree $(T,t)$ explains the full tuple $\mathcal{E}^*$, and thus
  $\mathcal{E}$ is \rsat.

  Finally, assume that $\mathcal{E}$ satisfies (S3).  By (S3.a.I),
  $G_{\Rone} \coloneqq (V,E_0\cup E_1\cup E_{\Rone})$ has $k>1$ strongly
  connected components $C_1, \dots, C_k$ collected in $\mathcal{C}$.
  Moreover, by (S3.a.II.i), there is a $C\in \mathcal{C}$ for which
  $G_0[C]$ is edge-less, where $C$ is $\ll$-minimal for some topological
  order $\ll$ on $G_{\Rone} /\{C_1,C_2, \dots, C_k\}$ and where
  $\mathcal{E}[\widehat C]$ is \rsat for $\widehat C =V\setminus C$.  Since
  $\mathcal{E}[\widehat C]$ is \rsat, $\mathcal{E}[\widehat C]$ can be
  extended to a full tuple $\mathcal{E}^*[\widehat C]=(E^*_0[\widehat C],
  E^*_1[\widehat C], E^*_{\Rone}[\widehat C])$ that is a explained by a
  Fitch-cotree $(\widehat T,\widehat t)$.  Moreover, $G_0[C]$ is edge-less
  and thus $\mathcal{E}[C] = (E_0[C],E_1[C],E_{\Rone}[C]) =
  (E_0[C],\emptyset,\emptyset)$.  We can now extend $\mathcal{E}[C]$ to a
  full tuple $\mathcal{E}^*[C] =(E^*_0[C],\emptyset,\emptyset)$ by adding
  all pairs $(x,y)\in \Cirr\setminus C$ to $E_0[C]$.  Clearly, $G^*_0[C]$
  remains edge-less. Hence, $\mathcal{E}^*[C]$ is explains by the
  Fitch-cotree $(T',t')$ where $T'$ is a star-tree whose root $\rho_{T'}$
  has label $t'(\rho_{T'})=0$. We finally join $(T',t')$ and
  $(\widehat{T},\widehat t)$ under a common root $\rho_T$ with to obtain
  the cotree $(T,t)$ where we put $t(\rho_T) = \Rone$ and assume that
  $\rho_{T'}$ is placed left of $\rho_{\widehat T}$.  The latter, in
  particular, ensures that $(T,t)$ is a Fitch-cotree.

  It remains to show that the full-set $\mathcal{E}^*$ explained by $(T,t)$
  extends $\mathcal{E}$. First observe that, by construction,
  $E_{\wr}[C]\subseteq E^*_{\wr}[C]$ and $E_{\wr}[\widehat C]\subseteq
  \ E^*_{\wr}[\widehat C]$ for $\wr\in \{0,1,\Rone\}$.  By construction of
  $(T,t)$, we have $\lca_T(x,y) = \rho_T$ for all $x\in C$ and $y\in
  \widehat C$.  By construction, $(T,t)$ explains $\mathcal{E}^*[C]$ and
  $\mathcal{E}^*[\widehat C]$ and thus, $\mathcal{E}^*[C]$ extends
  $\mathcal{E}[C]$ while and $\mathcal{E}^*[\widehat C]$ extends
  $\mathcal{E}[\widehat C]$.

  Hence, it remains to consider all vertices $x\in C$ and $y\in \widehat
  C$.  Let $x\in C$ and $y\in \widehat C$.  Since $C$ is $\ll$-minimal for
  some topological order $\ll$ on $H\coloneqq G_{\Rone} /\{C_1,C_2, \dots,
  C_k\}$, there is no edge $(C',C)$ in $H$ for any $C'\in \mathcal C$
  with $C'\subseteq \widehat C$. Hence, there is no edge $(y,x)$
  in $G_{\Rone}$.  Therefore, any edge between $x$ and $y$ that is
  contained in a relation of $\mathcal{E}$ must satisfy $(x,y)\in
  E_{\Rone}$ (since $E_0$ and $E_1$ are symmetric).  Since $\rho_{T'}$ is
  placed left of $\rho_{\widehat T}$, it follows that $x\in C$ is placed
  left of $y\in C'$ for every $C'\subseteq \widehat C$.  This, together
  with the fact that $\lca_T(x,y) = \rho_T$ and $t(\rho_T) = \Rone$,
  implies that $(x,y)\in E_{\Rone}$ is correctly explained for all $x\in C$
  and $y\in \widehat C$.  Moreover, for every $x\in C$ and $y\in \widehat
  C$, we have by construction of $(T,t)$ that $(x,y)\in E^*_{\Rone}$ and
  $(y,x)\notin E^*_{\Rone}$.  Hence, $E_{\Rone}\subseteq E^*_{\Rone}$
  holds.  Thus, the full-set $\mathcal{E}^*$ that is explained by $(T,t)$
  extends $\mathcal{E}$ and, therefore, $\mathcal{E}$ is $\rsat$.

  For the \textit{only-if} direction, we assume that the partial tuple
  $\mathcal{E} = (E_0,E_1,E_{\Rone})$ on $V$ is \rsat. If $|V|=1$, then
  (S1) is obviously satisfied. We therefore assume $|V|\geq 2$ from here
  on. Let $\mcEs= (E_0^*,E_1^*,E_{\Rone}^*)$ be a full tuple that is \rsat
  and that extends $\mathcal{E}$. By Cor.~\ref{cor:EquiDefSat}, there is a
  discriminating Fitch-cotree $(T,t)$ endowed with the sibling order order
  $<$ that explains $\mathcal{E}$ and $\mathcal{E}^*$ and has leaf set
  $V$. Moreover, the root $\rho$ of $T$ has at $r\geq 2$ children
  $v_1,\dots, v_r$ ordered from left to right according to $<$ and has one
  of the three labels $0,1,\Rone$. In the following, $L_i = \{x\in L(T)\mid
  x\preceq v_i\}$ denotes the set of all leaves $x$ of $T$ with $x\preceq
  v_i$. We make frequent use of the fact that $\lca(x,y)=\rho$ and thus,
  $t(\lca(x,y))=t(\rho)$ for all $x\in L_i, y\in L_j$ with $i\neq j$
  without further notice. We proceed by considering the three possible
  choices of $t(\rho)$.

  If $t(\rho)=0$, then Lemma \ref{lem:0=edge-less} implies that $G_0 =
  (V,E_1\cup E_{\Rone})$ is edge-less and thus $\mathcal{E}$ satisfies
  (S1).  If $t(\rho)=1$, then  $t(\rho) =t(\lca(x,y))=1$ for all
  $x\in L_i, y\in L_j$ with $i\neq j$. Hence, $G_1^* = (V,E^*_0\cup
  E_{\Rone}^*)$ must be disconnected. Since $E_0\subseteq E_0^*$ and
  $E_{\Rone}\subseteq E^*_{\Rone}$, $G_1 = (V,E_0\cup E_{\Rone})$ is a
  subgraph of $G^*_1$ and thus, $G_1$ is also disconnected. For each of the
  connected components $C$ of $G_1$, Lemma \ref{lem:hereditary} implies
  that $\mathcal{E}[C]$ is \rsat. Therefore, $\mathcal{E}$ satisfies (S2).

  If $t(\rho)=\Rone$ we have $t(\lca(x,y))=\Rone$ for all $x\in L_i$, $y\in
  L_j$ with $i\neq j$. Since $(T,t)$ explains $\mathcal{E}^*$, we observe
  that $(x,y)\in E^*_{\Rone}$ and $(y,x)\notin E^*_{\Rone}$ for all $x\in
  L_i$, $y\in L_j$ with $1\leq i< j\leq r$. Thus, $G^*_{\Rone} \coloneqq
  (V,E^*_0\cup E^*_1\cup E^*_{\Rone})= G^*_{\Rone}[L_1]\djoin\dots \djoin
  G^*_{\Rone}[L_r]$. Hence, $G^*_{\Rone}$ contains more than one strongly
  connected component (which may consist of a single vertex).  In
  particular, each strongly connected component of $G^*_{\Rone}$ must be
  entirely contained in some $L_i$, $i\in \{1,\dots,r\}$.  Let
  $\mathcal{C}$ be the set of strongly connected components of
  $G_{\Rone}$. Since $G_{\Rone}$ is a subgraph of $G^*_{\Rone}$,
  $G_{\Rone}$ contains more than one strongly connected component and
  therefore, $\mathcal{E}$ satisfies (S3.a.I). In particular, each $C\in
  \mathcal{C}$ must be contained in some strongly connected component of
  $G^*_{\Rone}$. Taken the latter arguments together, each $C\in
  \mathcal{C}$ must satisfy $C\subseteq L_i$ for some $i\in \{1,\dots,r\}$.
  
  Consider now the left-most child $v_1$ of $\rho$. Since $(T,t)$ is a
  discriminating Fitch-cotree, this vertex $v_1$ is either a leaf of $T$ or
  $t(v_1) = 0$.  In either case, $G_0[L_1]$ is edge-less
  (cf.\ Lemma~\ref{lem:0=edge-less}).  By the latter arguments and since
  $L_1\neq \emptyset$, there is some $C\in \mathcal{C}$ with $C\subseteq
  L_1$.  Since $G_0[C]\subseteq G_0[L_1]$ is edge-less, Condition
  (S3.a.II.i) holds.
  
  Let $C'\in \mathcal{C}\setminus\{C\}$. Suppose first that $C'\subseteq
  L_1$.  Assume, for contradiction, that there are edges $(x,y)$ or $(y,x)$
  with $x\in C$ and $y\in C'$ in $G_{\Rone}$. Since $G_0[L_1]$ is
  edge-less, we have $(x,y)\in E_0$ or $(y,x)\in E_0$. Since $E_0$ is
  symmetric $(x,y)\in E_0$ implies $(y,x)\in E_0$ and \emph{vice
  versa}. Hence, the union $C\cup C'$ would be strongly connected in
  $G_{\Rone}$; a contradiction since $C$ and $C'$ are distinct.  Hence,
  there cannot be any edge between vertices in $C$ and $C'$ in $G_{\Rone}$
  whenever $C'\subseteq L_1$.  Suppose now that $C'\subseteq L_i$, $1<i\leq
  r$. As argued above, $(x,y)\in E^*_{\Rone}$ and $(y,x)\notin E^*_{\Rone}$
  for all $x\in C \subseteq L_1 $ and $y\in C' \subseteq L_i$.  Since
  $E_{\Rone}\subseteq E^*_{\Rone}$, any to adjacent vertices $x\in C$ and
  $y\in C'$ in $G_{\Rone}$ must satisfy $(x,y)\in E_{\Rone}$ and
  $(y,x)\notin E_{\Rone}$.

  Let $\mathcal{C} = \{C_1,\dots,C_k\}$, $k>1$. Hence, the quotient $H =
  G_{\Rone} /\{C_1,C_2, \dots, C_k\}$ is a DAG.  By the latter arguments,
  there are no edges $(y,x)$ with $x\in C$ and $y\in C'$ for any $C'\in
  \mathcal{C}\setminus\{C\}$.  Hence, there is no edge $(C',C)$ in $H$ for
  any $C'\in \mathcal{C}\setminus\{C\}$. Since $H$ is a DAG, it
    admits a topological order
  $\ll$ such that $C$ is $\ll$-minimal. Hence, (S3.a.II.ii) is
  satisfied.  Finally, Lemma \ref{lem:hereditary} implies that
  $\mathcal{E}[V\setminus C]$ is \rsat and thus (S3.b) is satisfied.  In
  summary, $\mathcal{E}$ satisfied (S3).
\hfill\qed\end{proof}

\section{Recognition Algorithm and Computational Complexity}

\begin{figure}[ht]
  \centering
  \includegraphics[width = .9\textwidth]{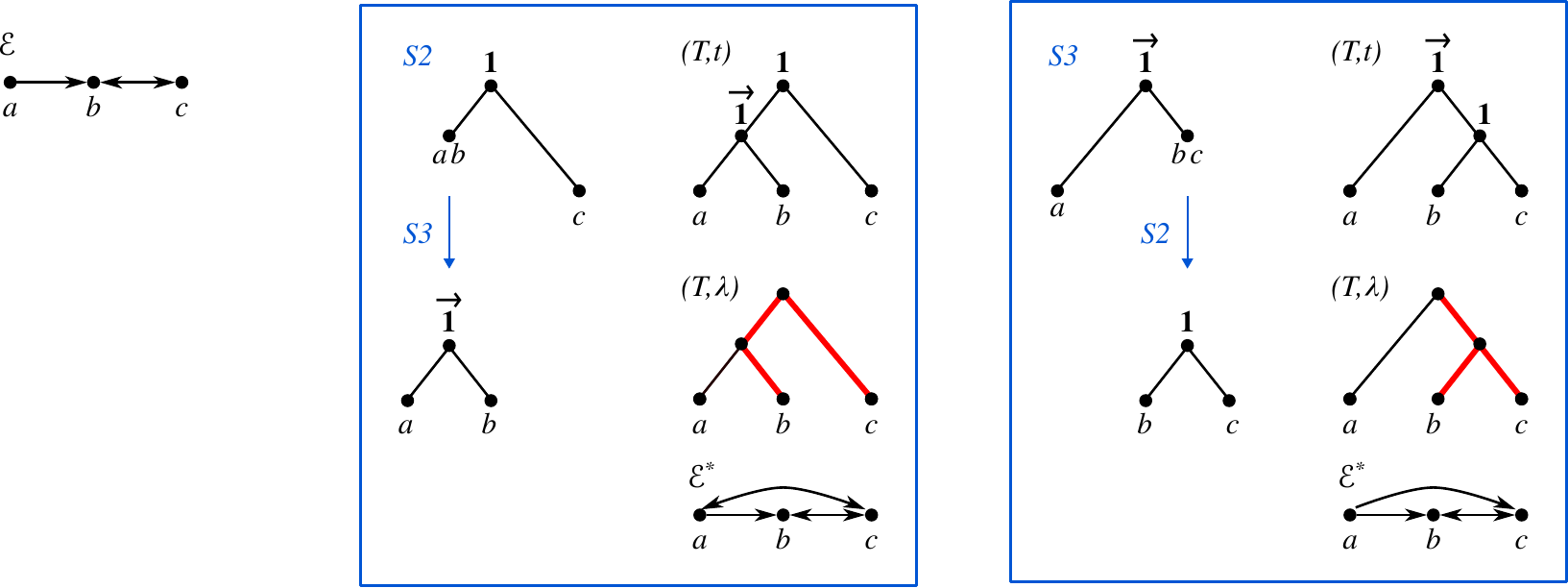}
  \caption{A partial tuple $\mathcal{E} = (E_0,E_1,E_{\protect \Rone})$ on
    $V=\{a,b,c\}$ with $E_0 = \emptyset$, $E_1 = \{(b,c),(c,b)\}$
    (bi-directional arc), and $E_{\protect \Rone} = \{(a,b)\}$ (single arc)
    is shown on the left.  Observe that (S1) is not satisfied while (S2.a)
    and (S3.a) hold for $\mathcal{E}$. Application of the different rules
    and subsequent construction of the Fitch-cotrees that explain the
    subgraphs induced by the respective (strongly) connected components
    results in two Fitch-cotrees that both explain $\mathcal{E}$. Hence, we
    obtain two different edge-labeled Fitch-trees $(T,\lambda)$ (with
    HGT-edges drawn in bold-red) that both explain $\mathcal E$.  }
\label{fig:simpleExample}
\end{figure}

The proof of Thm.~\ref{thm:Characterize-Partial} provides a recipe to
construct a Fitch-cotree $(T,t)$ explaining a tuple $\mathcal{E}$.  We
observe, furthermore, that two or even all three alternatives (S1), (S2.a),
and (S3.a) may be satisfied simultaneously, see
Fig.~\ref{fig:simpleExample} for an illustrative example. In this case, it
becomes necessary to check stepwisely whether conditions (S2.b) and/or
(S3.b) holds. Potentially, this result in exponential effort to determine
recursively whether $\mathcal{E}[C]$ or $\mathcal{E}[V\setminus C]$ is
\rsat. The following simple lemma shows, however, that the alternatives
always yield consistent results:
\begin{lemma}\label{lem:order-of-rules}
  Let $\mathcal{E} = (E_0,E_1,E_{\Rone})$ be a partial tuple on $V$. Then
  \begin{itemize}
  \item[] (S1) and (S2.a) implies (S2.b);
  \item[] (S1) and (S3.a) implies (S3.b);
  \item[] (S2a) and (S3a) implies that (S2.b) and (S3.b) are equivalent.
  \end{itemize}
\end{lemma}
\begin{proof}
  If (S1) holds, then Thm.~\ref{thm:Characterize-Partial} implies that
  $\mathcal{E}$ is \rsat. If (S2.a) holds, then heredity (Lemma
  \ref{lem:hereditary}) implies that $\mathcal{E}[C]$ is \rsat and thus
  (S2.b) is satisfied.  Analogously, if (S3.a) holds, then
  $\mathcal{E}[V\setminus C]$ is \rsat and thus (S3.b) holds. Now suppose
  (S2a) and (S3a) are satisfied but (S1) does not hold. Then $\mathcal{E}$
  is \rsat if and only if one of (S2.b) or (S3.b) holds; in the affirmative
  case, heredity again implies that both (S2.b) and (S3.b) are satisfied.
\hfill\qed\end{proof}
It follows that testing whether $\mathcal{E}$ can be achieved by checking
if any one of the three conditions (S1), (S2), or (S3) holds and, if
necessary, recursing down on $\mathcal{E}[C]$ or $\mathcal{E}[V\setminus
  C]$.  This give rise to Alg.~\ref{alg:build-cotree}.

\begin{algorithm}[htbp]
  \caption{Recognition of \rsat partial tuple $\mathcal{E}$ on $V$ and
    reconstruction of a cotree $(T,t)$ that explains $\mathcal{E}$.}
  \label{alg:build-cotree} 
  \begin{algorithmic}[1] \footnotesize
    \Require{Partial tuples $\mathcal{E} = (E_0, E_1, E_{\protect\Rone})$}
    \Ensure{A cotree $(T,t)$ that explains $\mathcal{E}$, if one exists or the statement   ``$\mathcal{E}$ is not Fitch-satisfiable''} \\
    \State Call \textsc{BuildFitchCotree}($V, \mathcal{E}$)
    \medskip
    \smallskip
    \Function{\textsc{BuildFitchCotree}}{$V, \mathcal{E}=
      (E_0, E_1, E_{\protect\Rone} )$}
      
    \Comment{$G_0$,  $G_1$ and $G_{\protect\Rone}$ are defined as in
      Thm.~\ref{thm:Characterize-Partial} for given $\mathcal{E}$}
    \If{$|V| = 1$} \Return{the cotree $((V, \emptyset), \emptyset)$}
    \label{lin:r0} 
    \ElsIf{$G_0$ is edge-less (\textbf{otherwise if} $G_1$  is disconnected)}
    \label{lin:r1} 
    
    \Comment{check (S1) (resp., (S2))}
    
    \State $\mathcal{C} \coloneqq $ the set of connected components
	   $\{ C_1, \dots, C_k\}$ of $G_0$ (resp. $G_1$)
    \State $\mathcal{T} \coloneqq $ set
    	$\{\textsc{BuildFitchCotree}(C_i,\ \mathcal{E}[C_i])
    		\mid C_i \in \mathcal{C}\}$\label{lin:run-tree1}
    \State \Return{the cotree from joining the cotrees in
    	  $\mathcal{T}$ under a new root labeled $0$ (resp. $1$)}
    \ElsIf{$G_{\protect\Rone}$ has more than one strongly connected
      component}
    \label{lin:r3}
    
        \Comment{check (S3)}
    \State $\mathcal{C} \coloneqq $ the set of strongly connected
    components $\{C_1, \dots, C_k\}$ of $G_{\protect\Rone}$
    \State $I\gets \emptyset$
    \ForAll{$i\in \{1,\dots,k\}$}
       \If{$G_0[C_i]$ is edge-less} \label{lin:I}
       $I\gets I\cup \{i\}$	
       \EndIf
    \EndFor
    
    \If{$I=\emptyset$} 
    \State Halt and output: ``$\mathcal{E}$ is not Fitch-satisfiable''
    \Comment{(S3.a.II.i) not satisfied}
    \ElsIf{there is no topological order $\ll$ of $G_{\Rone} /
      \{C_1, \dots, C_k\}$ with $\ll$-minimal element $C_i$ with $i\in I$}	 	
    \State Halt and output: ``$\mathcal{E}$ is not Fitch-satisfiable''
    \Comment{(S3.a.II.ii) not satisfied}
    \Else		 
    \State $\ll \coloneqq$ a topol.\ order on the quotient
    $G_{\Rone} / \{C_1, \dots, C_k\}$ with  $\ll$-minimal element
    $C^*\coloneqq C_i$ for some $i\in I$
    \State $\mathcal{T} \coloneqq
    \{\textsc{BuildFitchCotree}(C,\ \mathcal{E}[C])
    \mid C \in \{C^*, V\setminus C^* \}$
    \State \Return{the cotree $(T,t)$ obtained by joining the
      cotrees in $\mathcal{T}$	under a new root with label
      $\protect\Rone$, \hspace{.95cm}
      where the tree $(T^*,t^*)$ that explains $C^*$ is placed
      left of the tree $(\widehat T,\widehat t)$ that explains
      $V\setminus C$}
    \EndIf
    \label{lin:r3-end}
    \Else
    \State Halt and output: ``$\mathcal{E}$ is not Fitch-satisfiable''
    \EndIf
    \EndFunction
  \end{algorithmic}
\end{algorithm}

\begin{lemma}
  \label{lem:algcorr}
  Let $\mathcal{E} = (E_0, E_1, E_{\Rone} )$ be a partial tuple. Then
  Alg.~\ref{alg:build-cotree} either outputs a Fitch-cotree $(T,t)$
  that explains $\mathcal{E}$ or recognizes that $\mathcal{E}$ is not
  \rsat.
\end{lemma}
\begin{proof}
  Let $\mathcal{E}=(E_0,E_1,E_{\Rone})$ be a partial tuple on $V$ that serves
as input for Alg.\ \ref{alg:build-cotree}. In Alg.\ \ref{alg:build-cotree},
the function \textsc{BuildFitchCotree} is called and the algorithm we will
try and build a Fitch-cotree explaining a full \rsat tuple that extends
$\mathcal{E}$. By Lemma~\ref{lem:order-of-rules}, the order in which (S1),
(S2), and (S3) is tested is arbitrary.  If $|V| = 1$, then $\mathcal{E}$ is
already full and \rsat, and $(T,t) = ((V, \emptyset);\emptyset)$ is a valid
Fitch-cotree explaining $\mathcal{E}$, and thus returned on Line
\ref{lin:r0}. If none of the conditions (S1), (S2), or (S3) is satisfied,
Thm.~\ref{thm:Characterize-Partial} implies that $\mathcal{E}$ is not \rsat,
and Alg.~\ref{alg:build-cotree} correctly returns ``$\mathcal{E}$ is not
\rsat''.

If Rule (S1) or (S2.a), resp., is satisfied (Line \ref{lin:r1}), then
Alg.\ \ref{alg:build-cotree} is called recursively on each of the connected
components defined by $G_0$ or $G_1$, respectively.  In case of (S1), the
connected components are all single vertices and we obtain a star-tree
$(T,t)$ whose root is labeled ``0'' as outlined in the proof of the
\emph{if} direction of Thm.~\ref{thm:Characterize-Partial}. Since $(T,t)$
is a Fitch-cotree that explains $\mathcal{E}$, it follows that
$\mathcal{E}$ is \rsat. In case of (S2.a), condition (S2.b) must be
tested. If each of the connected components are indeed \rsat, the obtained
Fitch-cotrees are joined into a single cotree $(T,t)$ that explains a full
tuple $\mathcal{E}^*$ which extends $\mathcal{E}$. In particular, since no
constraints are imposed on the root $\rho_T$, which has label ``1'',
$(T,t)$ is a Fitch-cotree. Consequently, $\mathcal{E}$ is \rsat and
explained by $(T,t)$.

Suppose, finally, that $\mathcal{E}$ satisfies (S3.a) and let $\mathcal{C}
\coloneqq \{C_1, \dots, C_k\}$ be the the set of strongly connected
components of $G_{\protect\Rone}$. Since (S3.a.II) is satisfied, there must
be a $C_i\in \mathcal{C}$ for which $G[C_i]$ is edge-less, i.e., the set
$I$ as computed in Line \ref{lin:I} cannot be empty. In particular, there
is a topological order $\ll$ of $G_{\Rone} / \{C_1, \dots, C_k\}$ with
$\ll$-minimal element $C^*\coloneqq C_i$ for some $i\in I$. It is now
checked, if $\mathcal{E}[C^*]$ and $\mathcal{E}[\widehat C]$ are \rsat,
where $\widehat C \coloneqq V\setminus C$.  If this is the case, two
Fitch-cotrees $(T^*,t^*)$ and $(\widehat T, \widehat t)$ that explain
$\mathcal{E}[C^*]$ and $\mathcal{E}[\widehat C]$ are returned. We then join
these cotrees in under a new root with label $\Rone$, where the tree
$(T^*,t^*)$ is placed left from $(\widehat T, \widehat t)$. The latter
results in a cotree $(T,t)$ that explains $\mathcal{E}$. It remains to show
that $(T,t)$ is a Fitch-cotree. Since, $\rho_T$ has precisely two children
$\rho_{T^*}$ and $\rho_{\widehat T}$ and $\rho_{T^*}$ is left from
$\rho_{\widehat T}$, we must show that that the tree $(T^*,t^*)$ computed
in the recursive calls does not contain an inner vertex $v$ with label
$t(v)\neq 1$. The recursive call with $\mathcal{E}[C^*]$ first checks if
$|C^*|=1$ (line \ref{lin:r0}) and the continues with checking if $G_0[C^*]$
is edge-less (\ref{lin:r1}). Since $G_0[C^*]$ is edge-less, at least one of
the steps are satisfied and either the single-vertex tree or a star tree
whose root has label ``0'' with be returned. Consequently, $(T^*,t^*)$ does
not contain an inner vertex $v$ with label $t(v)\neq 1$ and thus, $(T,t)$
is a Fitch-cotree that explains $\mathcal{E}$. Hence, $\mathcal{E}$ is
\rsat.
\hfill\qed\end{proof}

\begin{theorem}
  Let $\mathcal{E} = (E_0, E_1, E_{\Rone} )$ be a partial tuple, $n = |V|$
  and $m =|E_0 \cup E_1 \cup E_{\Rone}|$.  Then,
  Alg.~\ref{alg:build-cotree} computes a Fitch-cotree $(T,t)$ that explains
  $\mathcal{E}$ or identifies that $\mathcal{E}$ is not \rsat in 
  $O(n^2 + nm)$ time.
  \label{thm:algo}
\end{theorem}
\begin{proof}
  Correctness is established in Lemma~\ref{lem:algcorr}. We first note that
  in each single call of \textsc{BuildFitchCotree}, all necessary di-graphs
  defined in (S1), (S2) and (S3) can be computed in $O(n+m)$
  time. Furthermore, each of the following tasks can be performed in
  $O(n+m)$ time: finding the (strongly) connected components of each
  digraph, construction of the quotient graphs, and finding the topological
  order on the quotient graph using Kahn's algorithm
  \cite{Kahn:62}. Moreover, the vertex end edge sets $V[C]$ and
  $\mathcal{E}[C]$ for the (strongly) connected components $C$ (or their
  unions) can constructed in $O(n+m)$ time by going through every element
  in $V$ and $\mathcal{E}$ and assigning each pair to their respective
  induced subset. Thus, every pass of \textsc{BuildFitchCotree} takes
  $O(n+m)$ time.  Since every call of \textsc{BuildFitchCotree} either
  halts or adds a vertex to the final constructed Fitch-cotree, and the
  number of vertices in this tree is bounded by the number $n$ of leaves,
  it follows that \textsc{BuildFitchCotree} is called at most $O(n)$ times
  resulting in an overall running time $O(n(n+m))$.
\hfill\qed\end{proof}

Instead of asking only for the existence of a Fitch-completion of a tuple
$\mathcal{E}=(E_0,E_1, E_{\Rone})$, it is of interest to ask for the
completion that maximizes a total score for the the pairs of distinct
vertices $x,y$ that are not already classified by $\mathcal{E}$, i.e.,
$\{x,y\} \in \overline{\mathcal{E}} \coloneqq \{ \{x,y\} \notin (E_0\cup
E_1\cup E_{\Rone})^{sym}\}$.  For every pair of vertices there are four
possibilities $x\btwn y \in \{x\rightleftharpoons y,x\to y, x\leftarrow y,
x\noedge y \}$. The score $w(x\btwn y)$ may be a log-odds ratio
for observing one of the four possible xenology relationship as determined
from experimental data. Let us write $F=F_{\mathcal{E}^*}$ for the Fitch
graph defined by the extension $\mathcal{E}^*$ of $\mathcal{E}$ and associate with it
the the total weight of relations added, i.e.,
\begin{equation}
  f(F) = \sum_{\{x,y\}\in\overline{\mathcal{E}}} w(F[\{x,y\}]) 
\end{equation}

The weighted Fitch-graph completion problem can also be seen as special
case of the problem with empty tuple
$\mathcal{E}^{\emptyset}\coloneqq(\emptyset,\emptyset,\emptyset)$.  To see
this, suppose first that an arbitrary partial input tuple $\mathcal{E}$ is
given.  For each two vertices $x,y$ for which
$\{x,y\}\notin\overline{\mathcal{E}}$ the induced graphs $F[\{x,y\}]$ is
well-defined and we extend the weight function to all pairs of vertices by
setting, for all input pairs, $w(F[\{x,y\}])=m_0$ and $w(x\btwn y)=-m_0$
for $(x\btwn y)\ne F[\{x,y\}]$, where $m_0\gg
|V|^2\max_{\btwn,\{x,y\}\in\overline{\mathcal{E}}} |w(x\btwn y)|$.  Now
consider the weighted Fitch graph completion problem with this weight
function and an empty tuple $\mathcal{E}^{\emptyset}$. The choice of
weights ensures that any Fitch graph $F'$ maximizing $f(F')$ induces
$F'[\{x,y\}]=F[\{x,y\}]$ for all pairs $\{x,y\}$ in the input tuple,
because not choosing $F[\{x,y\}]$ reduces the score by $2m_0$ while the
total score of all pairs not specified in the input is smaller than $m_0$.
In order to study the complexity of this task, it therefore suffices to
consider the following decision problem:

\begin{problem}[\PROBLEM{Fitch Completion Problem (FC)}]\ \\
  \begin{tabular}{ll}
    \emph{Input:}    & A set $V$, an assignment of 
                       four weights $w_{xy}(x\btwn y)$ to all distinct \\
                     &  $x,y\in V$ where $\btwn \in
                       \{\rightleftharpoons,\to,
                       \leftarrow, \noedge  \}$,  
		      and an integer $k$.\\
    \emph{Question:} & Is there a Fitch graph $F = (V,E)$ such that\\
                     &  $f(F) = \sum_{\substack{x,y\in V\\ x\ne y}}
                        w_{xy}(F[\{x,y\}])\geq k$?
  \end{tabular}
\end{problem}

\noindent
For the NP-hardness reduction, we use the following NP-complete problem
\cite{Karp1972}
\begin{problem}[\PROBLEM{Maximum Acyclic Subgraph Problem (MAS)}]\ \\
  \begin{tabular}{ll}G
    \emph{Input:}    & A digraph $G =(V,E)$    and an integer $k$.\\
    \emph{Question:} & Is there a subset $E'\subseteq E$ such that $|E'|\geq
                       k$ and $(V,E')$ is a DAG?
  \end{tabular}
\end{problem}

\begin{theorem}\label{thm:FC-NPc}
  \PROBLEM{FC} is NP-complete.
\end{theorem}
\begin{proof}
  We claim that \PROBLEM{FC} is in NP. To see this, let $F=(V,E)$ be a
  given digraph. We can check whether $f(F) \geq k$ in polynomial time by
  iterating over all edges in $F$. In addition, by
  Thm.~\ref{thm:CharFitch}, we can check whether $F$ is a Fitch graph in
  polynomial time by iterating over all 3-subsets of $V$ and verifying that
  none of the induced a forbidden subgraph of Fitch graphs.

  To prove NP-hardness, let $(G =(V,E), k)$ be an instance of
  \PROBLEM{MAS}. We take as input for \PROBLEM{FC} the set $V$, the integer
  $k$ and the following weights for all distinct $x,y\in V$:
  \begin{itemize}
  \item[(i)] If $(x,y)\in E$, then put $w_{xy}(x\to y)=1$  	
  \item[(ii)] If $(x,y)\notin E$, then put $w_{xy}(x\to y)=0$
  \item[(iii)] Put $w_{xy}(x  \noedge y) = 0$ and
    $w_{xy}(x  \rightleftharpoons y) = -|V|^2$. 
  \end{itemize}
  Note that Condition (i) ensures that, for all $x,y\in V$, we have
  $w_{xy}(y\to x)=1$ if $(y,x)\in E$ and $w_{xy}(x\to y)= w_{xy}(y\to x)
  =1$ whenever both $(x,y)$ and $(y,x)$ are edges in $G$.

  Suppose first that there is a subset $E'\subseteq E$ such that $|E'|\geq
  k$ and $G'=(V,E')$ is a DAG. Hence, for any $x,y\in V$ not both $(x,y)$
  and $(y,x)$ can be contained in $E'$. This, together with the
  construction of the weights implies that $f(G') = |E'|\geq k$. We now
  extend $G'$ to a Fitch graph $F$. To this end, observe that $G'$ admits a
  topological order $\ll$. We now add for all pairs $x,y$ with $x \ll y$
  and $(x,y)\notin E'$ the edge $(x,y)$ to obtain the di-graph
  $F$. Clearly, $\ll$ remains a topological order of $F$ and, therefore,
  $F$ is a DAG. This with the fact that $F$ does not contain bi-directional
  edges or non-adjacent vertices together with Cor.\ \ref{cor:DAG-fitch}
  implies that $F$ is a Fitch graph.  In particular, $f(F)\geq f(G')\geq
  k$.

  Assume now that there is a Fitch graph $F$ such that $f(F)\geq k$.  Since
  $w_{xy}(x \rightleftharpoons y) = -|V|^2$ and the weight for any
  uni-directed edge and every pair of non-adjacent vertices is $0$ or $1$
  and the maximum number of edges in $F$ is $2\cdot\binom{|V|}{2} = |V|^2-
  |V|<|V|^2$, $f(F)\geq k$ implies that $F$ cannot contain bidirectional
  edges. By Cor.\ \ref{cor:acyclicFitch}, $F$ is acyclic. Now, take the
  subgraph $G'$ of $F$ that consists of all edges with weight $1$. Clearly,
  $G'$ remains acyclic and $f(F)=f(G')\geq k$. By construction of the
  weights, all edges of $G'$ must have been contained in $G$ and thus,
  $G'\subseteq G$ is an acyclic subgraph of $G$ containing at least $k$
  edges.
\hfill\qed\end{proof}


%
%


%
 \bibliographystyle{splncs04}
 \bibliography{fitchopt}

\begin{thebibliography}{10}
\providecommand{\url}[1]{\texttt{#1}}
\providecommand{\urlprefix}{URL }
\providecommand{\doi}[1]{https://doi.org/#1}

\bibitem{Boecker:98}
B{\"o}cker, S., Dress, A.W.M.: Recovering symbolically dated, rooted trees from
  symbolic ultrametrics. Adv. Math.  \textbf{138},  105--125 (1998).
  \doi{10.1006/aima.1998.1743}

\bibitem{Corneil:81}
Corneil, D.G., Lerchs, H., Steward~Burlingham, L.: Complement reducible graphs.
  Discr. Appl. Math.  \textbf{3},  163--174 (1981)

\bibitem{Corneil:85}
Corneil, D.G., Perl, Y., Stewart, L.K.: A linear recognition algorithm for
  cographs. SIAM J. Computing  \textbf{14},  926--934 (1985)

\bibitem{CP-06}
Crespelle, C., Paul, C.: Fully dynamic recognition algorithm and certificate
  for directed cographs. Discr. Appl. Math.  \textbf{154},  1722--1741 (2006)

\bibitem{EHPR:96}
Engelfriet, J., Harju, T., Proskurowski, A., Rozenberg, G.: Characterization
  and complexity of uniformly nonprimitive labeled 2-structures. Theor. Comp.
  Sci.  \textbf{154},  247--282 (1996)

\bibitem{Geiss:18a}
Gei{\ss}, M., Anders, J., Stadler, P.F., Wieseke, N., Hellmuth, M.:
  Reconstructing gene trees from {Fitch}'s xenology relation. J. Math. Biol.
  \textbf{77},  1459--1491 (2018). \doi{10.1007/s00285-018-1260-8}

\bibitem{gurski2017dynamic}
Gurski, F.: Dynamic programming algorithms on directed cographs. Statistics,
  Optimization \& Information Computing  \textbf{5}(1),  35--44 (2017).
  \doi{10.19139/soic.v5i1.260}

\bibitem{hellmuth2023resolving}
Hellmuth, M., Scholz, G.E.: Resolving prime modules: The structure of
  pseudo-cographs and galled-tree explainable graphs. Tech. rep., arXiv (2023).
  \doi{10.48550/arXiv.2211.16854}

\bibitem{HS:19}
Hellmuth, M., Seemann, C.R.: Alternative characterizations of {Fitch}'s
  xenology relation. J. Math. Biol.  \textbf{79}(3),  969--986 (2019).
  \doi{10.1007/s00285-019-01384-x}

\bibitem{Kahn:62}
Kahn, A.B.: Topological sorting of large networks. Communications of the ACM
  \textbf{5},  558--562 (1962). \doi{10.1145/368996.369025}

\bibitem{Karp1972}
Karp, R.M.: Reducibility among combinatorial problems. In: Miller, R.E.,
  Thatcher, J.W., Bohlinger, J.D. (eds.) Complexity of Computer Computations:
  Proceedings of a symposium on the Complexity of Computer Computations, pp.
  85--103. Springer US, Boston, MA (1972). \doi{10.1007/978-1-4684-2001-2_9}

\bibitem{mcconnell2005linear}
McConnell, R.M., De~Montgolfier, F.: Linear-time modular decomposition of
  directed graphs. Discr. Appl. Math.  \textbf{145}(2),  198--209 (2005).
  \doi{10.1016/j.dam.2004.02.017}

\bibitem{NEMWH:18}
N{\o}jgaard, N., El-Mabrouk, N., Merkle, D., Wieseke, N., Hellmuth, M.: Partial
  homology relations -- satisfiability in terms of di-cographs. In: Wang, L.,
  Zhu, D. (eds.) Computing and Combinatorics. Lecture Notes Comp. Sci., vol.
  10976, pp. 403--415. Springer International Publishing, Cham (2018).
  \doi{10.1007/978-3-319-94776-1_34}

\bibitem{Ravenhall:15}
Ravenhall, M., {\v{S}}kunca, N., Lassalle, F., Dessimoz, C.: Inferring
  horizontal gene transfer. PLoS Comp. Biol.  \textbf{11},  e1004095 (2015).
  \doi{10.1371/journal.pcbi.1004095}

\end{thebibliography}

\end{document}